\documentstyle[preprint,aps]{revtex}
\def\be{\begin{equation}}
\def\ee{\end{equation}}
\def\bea{\begin{eqnarray}}
\def\eea{\end{eqnarray}}
\begin{document}
\draft 
\title{Multiple-Time Higher-Order Perturbation Analysis \\ 
of the Regularized Long-Wavelength Equation}
\author{R. A. Kraenkel$^1$, M. A. Manna$^2$, V. Merle$^2$, 
J. C. Montero$^1$ and J. G. Pereira$^1$}
\vskip 0.5cm
\address{$^1$Instituto de F\'{\i}sica Te\'orica\\
Universidade Estadual Paulista\\
Rua Pamplona 145\\
01405-900\, S\~ao Paulo -- Brazil \\
\vskip 0.5cm
$^2$Physique Math\'ematique et Th\'eorique, URA-CNRS 768\\
Universit\'e de Montpellier II\\
34095 Montpellier Cedex 05 -- France}
%\date{today}
\maketitle
\begin{abstract}
By considering the long-wave limit of the regularized long 
wave (RLW) equation, we study its multiple-time higher-order 
evolution equations. As a first result, the equations of 
the Korteweg-de Vries hierarchy are shown to play a crucial
role in providing a secularity-free perturbation theory in 
the specific case of a solitary-wave solution. Then, as a 
consequence, we show that the related perturbative 
series can be summed and gives exactly the solitary-wave 
solution of the RLW equation. 
Finally, some comments and considerations are made on the 
N-soliton solution, as well as
on the limitations of applicability of the multiple scale
method in obtaining uniform perturbative series.
\end{abstract}
\vfill \eject

\section{Introduction}

The regularized long wave (RLW) equation,
\be
u_t + u_x - u_{xxt} - 6 u u_x = 0 \, ,
\label{bbm}
\ee
also known as Peregrine~\cite{pere} or Benjamin-Bona-
Mahony~\cite{bbm} equation, was originally introduced as an 
alternative for the Korteweg-de Vries (KdV) equation
\be
u_t + u_x + u_{xxx} - 6 u u_x = 0 \, .
\label{kdvu}
\ee
Despite having quite different dispersion properties, these two 
equations present some intimate relationship. For example, the 
linear dispersion relation of the RLW equation is
\be
\omega(k) = \frac{k}{1 + k^2} \, ,
\label{dr}
\ee
which, by the way, is the same as the dispersion relation of the 
shallow water wave equation~\cite{akns}. For long waves, $k$ is 
small and $\omega(k)$ can be expanded according to
\be
\omega(k) = k - k^3 + {\cal O}(k^5) \, .
\ee
The first two terms of this expansion coincide exactly with the 
complete linear dispersion relation of the KdV equation. Thus, 
for sufficiently long waves, the traveling-wave
solutions of Eqs.(\ref{bbm}) and (\ref{kdvu}) are expected to 
be quite similar. Despite of this, there is a deep difference 
between these two cases since a polynomial is definitely not 
equivalent to an infinite series. This difference, which appears 
when higher order terms of the dispersion relation expansion are 
considered, might also show up at higher order approximations in 
a perturbation theory.

As is well known, the KdV equation appears as governing the first 
relevant order of an asymptotic perturbation expansion describing 
weakly nonlinear dispersive waves. However, to make sense of it as 
really governing such waves, the large time behaviour of the 
perturbative series must be analyzed~\cite{koda1}. In other words, 
one needs to study the evolution equations of the higher 
order terms of the perturbative expansion to check for the existence 
or not of secular producing terms. This study, usually neglected 
in the derivation of the KdV equation, is essential to guarantee 
the uniformity of the perturbative expansion, rendering thus a 
real meaning to the KdV equation.

Motivated by the above considerations, we are going in this paper 
to apply a multiple time version~\cite{nois1} of the reductive 
perturbation method to study long waves as governed by the RLW 
equation. As we are going to see, the KdV equation appears at 
the lowest relevant order of the perturbative scheme. Then, by 
assuming a solitary-wave solution for the KdV equation, we consider 
the higher order approximations and we show that the corresponding 
solitary-wave related secular producing terms can be eliminated 
from every order of the perturbative scheme. The equations of 
the KdV hierarchy, which appear as a consequence of natural 
compatibility conditions, are shown to play a crucial role in 
the process of eliminating the secular producing terms. Once a 
secularity-free perturbative series is obtained, we show that it 
may be summed to give the exact solitary-wave solution of the RLW 
equation. We then close the paper with a discussion on the N-soliton 
case, as well as on the limitations of the multiple scale method.

\section{Multiple-Time Formalism}

To study the long-wave limit of the RLW equation we put
\be
k = \epsilon \kappa \, ,
\label{lw}
\ee
with $\epsilon$ a small parameter. In this limit, the dispersion 
relation (\ref{dr}) can be expanded as
\be
\omega(\kappa) = \epsilon \kappa - \epsilon^3 \kappa^3 + \epsilon^5 
\kappa^5 - \epsilon^7 \kappa^7 + \cdots \; .
\ee
Accordingly, the solution of the corresponding linear RLW equation 
can be written in the form
\be
u = a \exp i[k x - \omega(k) t] \equiv a \exp i \left[ \kappa 
\epsilon (x - t) + \epsilon^3 \kappa^3 t 
- \epsilon^5 \kappa^5 t + \epsilon^7 \kappa^7 t + \cdots \right] \, ,
\ee
where $a$ is a constant. As given by this solution, we define now a 
slow space
\be
\xi = \epsilon (x - t) \, ,
\label{ss}
\ee
as well as an infinity of properly normalized slow time variables:
\be
\tau_3 = \epsilon^3 t \quad ; \quad \tau_5 = - \epsilon^5 t \quad ; 
\quad \tau_7 = \epsilon^7 t \quad ; \quad {\rm etc}.
\label{st}
\ee
Consequently, we have
\be
\frac{\partial}{\partial x} = \epsilon \frac{\partial}{\partial \xi} 
\, ,
\label{dx}
\ee
and
\be
\frac{\partial}{\partial t} = - \epsilon \frac{\partial}{\partial \xi}
+ \epsilon^3 \frac{\partial}{\partial \tau_3} -
\epsilon^5 \frac{\partial}{\partial \tau_5} +
\epsilon^7 \frac{\partial}{\partial \tau_7} - \cdots \, .
\label{dt}
\ee
It is important to notice that the introduction of slow time variables 
normalized according to the dispersion relation expansion are such 
that they allow for an automatic elimination of the solitary-wave 
related secular--producing terms appearing in the evolution equations 
for the higher order terms of the wave--field~\cite{lecce}.

\section{Perturbation Theory}

The perturbative scheme consists of making the expansion
\be
u \equiv \epsilon^2 {\hat u} = \epsilon^2 (u_0 + \epsilon^2 u_2 + 
\epsilon^4 u_4 + \cdots) 
\, ,
\label{uhat}
\ee
and substituting it, together with Eqs.(\ref{dx}) and (\ref{dt}), 
into the RLW equation (\ref{bbm}). The result is the multiple--time 
equation
\bea
\left(\epsilon^3 \frac{\partial}{\partial \tau_3} - \epsilon^5 
\frac{\partial}{\partial \tau_5} + \cdots \right) {\hat u}
- \epsilon^2 \frac{\partial^2}{\partial \xi^2} \left(- \epsilon 
\frac{\partial}{\partial \xi} + 
\epsilon^3 \frac{\partial}{\partial \tau_3} - \epsilon^5 
\frac{\partial}{\partial \tau_5} + 
\cdots \right) {\hat u}&{}& 
\nonumber \\
- 3 \epsilon^3 \frac{\partial}{\partial \xi} \Big({u_0}^2 + 2 
\epsilon^2 u_0 u_2 + \cdots \Big) &=& 0 \, .
\label{ebbm}
\eea 
We proceed then to an order-by-order analysis of this equation.

At the lowest order, we obtain
\be
u_{0 \tau_3} = F_3 \equiv - u_{0 \xi\xi\xi} + 6 u_0 u_{0 \xi} = 0 \, ,
\label{kdv}
\ee
which is the KdV equation. Introducing an operator $L$, whose 
action on any component $u_n$ is given by the linearized KdV 
operator
\be
L u_n \equiv u_{n \tau_3} + u_{n \xi\xi\xi} - 6 (u_0 u_n)_{\xi} \, ,
\label{l}
\ee
the KdV equation (\ref{kdv}) can be rewritten in the form
\be
L u_0 = - 6 u_0 u_{0\xi} \, .
\label{lkdv0}
\ee
Our interest in this paper is concerned to solitary-waves. Thus 
we assume $u_0$ to be the solitary-wave 
solution of the KdV equation (\ref{kdv}),
\be
u_0 = - 2 \, \kappa^2 \, {\rm sech}^2 \, \theta_3 \, ,
\label{sw3}
\ee
where $\theta_3 = \kappa [\xi - 4 \kappa^2 \tau_3]$. In this case, 
Eq.(\ref{lkdv0}) becomes
\be
L u_0 = 48 \, \kappa^5 \, {\rm sech}^4 \, \theta_3 \, \tanh \, 
\theta_3 \, .
\label{lu0}
\ee

In the next order, we obtain the equation
\be
L u_2 = u_{0\tau_{3}\xi\xi} + u_{0\tau_5} \, .
\label{lu2}
\ee
The evolution of $u_0$ in the time $\tau_3$ is given by the 
KdV equation (\ref{kdv}), but the evolution of $u_0$ in the 
time $\tau_5$ is not known up to this point. However, the 
multiple--time formalism introduces constraints which determine 
uniquely the evolution of $u_0$ in any higher--order 
time~\cite{nois1}. To see how this is possible, let us make the 
following considerations.

First, to have a well ordered perturbative scheme, we impose 
that each one of the equations describing 
the higher-order times evolution of $u_0$ be $\epsilon$-independent 
when passing from the slow $(\kappa, u_0, \xi, \tau_{2n+1})$ to 
the laboratory coordinates $(k, u, x, t)$. This will select all 
possible terms to appear in $u_{0\tau_{2n+1}}$. For instance, 
the evolution of $u_0$ in the time $\tau_5$ 
is restricted to be of the form
\be
u_{0\tau_5} = \alpha_5 u_{0(5\xi )} + \beta_5 u_0 u_{0\xi \xi \xi } 
+ \gamma_5 u_{0\xi } u_{0\xi \xi } + \delta_5 u_0^2 u_{0\xi } \, ,
\ee
where $\alpha_5$, $\beta_5$, $\gamma_5$ and $\delta_5$ are 
constants to be determined. Then, by imposing the natural 
(in the multiple time formalism) compatibility condition
\be
\Big(u_{0\tau_3} \Big)_{\tau_{5}} = \Big(u_{0\tau_{5}} 
\Big)_{\tau_3} \, ,
\label{5compa}
\ee
it is possible to determine the above constants in terms of 
$\alpha_5$, which is left as a free-parameter. As it can be 
verified~\cite{nois1}, the resulting equation is the $5th$ 
order equation of the KdV hierarchy:
\be
u_{0\tau_5} = F_5 \equiv u_{0(5\xi)} - 10 u_0 u_{0\xi\xi\xi} 
- 20 u_{0\xi} u_{0\xi\xi} + 30 {u_0}^2 u_{0\xi} \, .
\label{kdv5}
\ee
The right-hand side of this equation would in principle appear 
multiplied by the free parameter $\alpha_5$, which would 
account for different possible normalizations of the 
time $\tau_5$. However, since we have already defined the 
slow time normalizations, this parameter was taken to be 
$1$ in order to have an agreement with the normalizations 
introduced in Eq.(\ref{st}). This is an important point since, 
as we have already said, it allows for an automatic elimination 
of the solitary-wave related secular producing terms appearing 
in the right--hand side of Eq.(\ref{lu2}). These terms, when 
$u_0$ is assumed to be a solitary--wave of the KdV equation, 
are always of the form~\cite{kota}
\be
u_{0[(2n+1)\xi]} \, ; \quad n = 0, 1, 2, \dots \, \, .
\ee
Thus, using Eqs.(\ref{kdv}) and (\ref{kdv5}) to describe 
respectively $u_{0\tau_3}$ and $u_{0\tau_5}$, Eq.(\ref{lu2}) 
becomes
\be
L u_2 = - 2 u_{0\xi} u_{0\xi\xi} - 4 u_{0} u_{0\xi\xi\xi} + 
30 {u_0}^2 u_{0\xi} \, .
\label{llu2}
\ee
We notice in passing that the substitution of Eqs.(\ref{kdv}) 
and (\ref{kdv5}), respectively for $u_{0\tau_3}$ and 
$u_{0\tau_5}$, with the properly normalized slow times allowed 
for an automatic elimination of all solitary-wave related 
secular producing terms of Eq.(\ref{lu2}). In fact, Eq.(\ref{llu2}) 
does not present any secular-producing term anymore. Moreover, 
we see that at this order $u_0$ must satisfy simultaneously the 
first two equations of the KdV hierarchy, respectively in the 
slow-times $\tau_3$ and $\tau_5$. Introducing the general definition
\be
\theta_{2n+1} = \kappa \left[ \xi - 4 \kappa^2 \tau_3 + 16 \kappa^4 
\tau_5 - \cdots + (-1)^{n} (2 \kappa)^{2n} \tau_{2n+1} \right] \, ,
\label{theta}
\ee
such a solitary-wave is given by
\be
u_0 = - 2 \, \kappa^2 \, {\rm sech}^2 \, \theta_5 \, ,
\label{sw5}
\ee
and Eq.(\ref{llu2}) becomes
\be
L u_2 = 192 \, \kappa^7 \, {\rm sech}^4 \, \theta_5 \, \tanh \, 
\theta_5 \, .
\label{lllu2}
\ee
Assuming a vanishing solution for the associated homogeneous 
equation, we can write the solution of 
this equation in the form
\be
u_2 = 4 \kappa^2 u_0 \, ,
\label{u2}
\ee
with $u_0$ given by (\ref{sw5}).

We proceed then to the next order, where we get
\be
L u_4 = - u_{0\tau_7} - u_{0\tau_5 \xi\xi} + u_{2\tau_5} + 
u_{2\tau_3 \xi\xi} + 6 u_2 u_{2\xi} \, .
\label{lu4}
\ee
Following the same scheme used before, we can use the 
compatibility condition 
\be
\Big(u_{0\tau_3} \Big)_{\tau_{7}} = \Big(u_{0\tau_{7}} 
\Big)_{\tau_3}
\label{7compa}
\ee
to obtain the evolution of $u_0$ in the time $\tau_7$. 
It is given by
\begin{eqnarray}
u_{0\tau_7} = F_7 \equiv &-& u_{0(7\xi)} + 14 u_0 u_{0(5\xi)} 
+ 42 u_{0\xi} u_{0(4\xi)} + 140 (v_0)^3 v_{0\xi} \nonumber \\ 
&+& 70 u_{0\xi\xi} u_{0\xi\xi\xi} - 280 u_0 u_{0\xi} u_{0\xi\xi} 
- 70(u_{0\xi})^3 - 70 {u_0}^2 u_{0\xi\xi\xi} \, ,
\label{kdv7}
\end{eqnarray}
which is exactly the $7th$ order equation of the KdV hierarchy. 
At this order, therefore, the solitary-wave must satisfy 
simultaneously the first three equations of the KdV hierarchy, 
respectively in the times $\tau_3$, $\tau_5$ and $\tau_7$. 
This means that   
\be
u_0 = - 2 \kappa^2 {\rm sech}^2 \, \theta_7 \, .
\label{sw7}
\ee
Now, by using Eq.(\ref{u2}) to express $u_2$, and the equations 
of the KdV hierarchy to express $u_{0\tau_7}$, $u_{0\tau_5}$ and 
$u_{0\tau_3}$, all secular producing terms of Eq.(\ref{lu4}) are 
automatically eliminated. Then, substituting the solution 
(\ref{sw7}), Eq.(\ref{lu4}) becomes
\be
L u_4 = 768 \, \kappa^9 \, {\rm sech}^4 \, \theta_7 \, \tanh \, 
\theta_7 \, .
\label{llu4}
\ee
Again, by assuming a vanishing solution for the associated 
homogeneous equation, the solution of this equation can be written 
as
\be
u_4 = (4 \kappa^2)^2 u_0 \, ,
\label{u4}
\ee
with $u_0$ given now by Eq.(\ref{sw7}).

This procedure can be repeated up to any higher order. In other 
words, we can use the compatibility condition
\be
\Big(u_{0\tau_3} \Big)_{\tau_{2n+1}} = \Big(u_{0\tau_{2n+1}} 
\Big)_{\tau_3}
\label{ncompa}
\ee
to obtain the evolution of $u_0$ in the time $\tau_{2n+1}$, 
which will turn out to be the $(2n+1)th$ equation of the 
KdV hierarchy. In this case, $u_0$ will represent a solitary-wave 
satisfying simultaneously the first $n$ equations of the KdV 
hierarchy:
\be
u_0 = - 2 \kappa^2 \, {\rm sech}^2 \, \theta_{2n+1} \, .
\label{swn}
\ee
The resulting secularity-free evolution equation at this order 
will be
\be
L \, u_{2n} = 3 \, (4)^{n+2} \, (\kappa)^{2n+5} \, {\rm sech}^4 \, 
\theta_{2n+3} \, \tanh \, \theta_{2n+3} \, .
\label{lun}
\ee
Assuming a vanishing solution for the associated homogeneous 
equation, the solution to this equation can be written in the form
\be
u_{2n} = (4 \kappa^2)^n \, u_0 \, ,
\label{u2n}
\ee
with $u_0$ given by Eq.(\ref{swn}). Extending this procedure {\it 
ad infinitum}, $u_0$ will represent a solitary-wave satisfying 
simultaneously all equations of the KdV hierarchy, and we obtain 
an exact solution for the RLW equation.

\section{Returning to the Laboratory Coordinates}

Let us take the solutions $u_{2n}$ and substitute them in the 
expansion (\ref{uhat}). Putting $u_0$ in evidence, we get
\be
u = \epsilon^2 u_0 \left[1 + 4 \epsilon^2 \kappa^2 + 16 
\epsilon^4 \kappa^4 + 64 \epsilon^6 \kappa^6 + \cdots \right] \, .
\ee
Now, the above series can be summed:
\be
1 + 4 \epsilon^2 \kappa^2 + 16 \epsilon^4 \kappa^4 + 64 
\epsilon^6 \kappa^6 + \cdots = 
\frac{1}{1 - 4 \epsilon^2 \kappa^2} \, .
\label{s}
\ee
Therefore, we get the RLW exact solution
\be
u = - \frac{2 \epsilon^2 \kappa^2}{1 - 4 \epsilon^2 \kappa^2} \, 
{\rm sech}^2 \left[\kappa \xi - 4 \kappa^3 \tau_3 + 16 \kappa^5 
\tau_5 - 64 \kappa^7 \tau_7 + \cdots \right] \, .
\ee
Then, by using Eqs.(\ref{lw}), (\ref{ss}) and (\ref{st}), we 
can rewrite $u$ in terms of the laboratory coordinates $(k, x, t)$. 
The result is
\be
u = - \frac{2 k^2}{1 - 4 k^2} \, {\rm sech}^2 \left[k x - k 
\left(1 + 4 k^2 + 16 k^4 + 64 k^6 + \cdots \right) \, t \right] \, .
\ee
Using again Eq.(\ref{s}) with $\epsilon \kappa=k$, we get finally
\be
u = - a \; {\rm sech}^2 \left[k \left(x - \frac{t}{1 - 4 k^2} \right) 
\right] \quad ; \quad a = \frac{2 k^2}{1 - 4 k^2} \; ,
\label{bbmsw}
\ee
which is the solitary-wave solution of the RLW equation (\ref{bbm}).

The RLW equation has another solution, given by
\be
u = b \; {\rm tanh}^2 \left[k \left(x - \frac{t}{1 + 8 k^2} \right) 
\right] \quad ; \quad b = \frac{2 k^2}{1 + 8 k^2} \; .
\label{bbmtan}
\ee
In fact, it is easy to see that Eq.(\ref{bbm}) is invariant 
under the transformation
\be
t^{\prime} = a^{-1} t \quad ; \quad x^{\prime} = x \quad ; 
\quad u^{\prime} = b - a u \; ,
\ee
where, if $u$ is given by Eq.(\ref{bbmsw}), $u^{\prime}$ turns 
out to be the solution given by Eq.(\ref{bbmtan}). By following 
the same procedure used to obtain the RLW solitary-wave solution 
(\ref{bbmsw}), it is also possible to use the multiple-time 
perturbative scheme to obtain the solution (\ref{bbmtan}). 
This is done by choosing 
\be
u_0 = 2 \kappa^2 {\rm tanh}^2 \left(\kappa \xi - 8 \kappa^3 
\tau_3 \right) \; ,
\ee
instead of (\ref{sw3}) as the solution for the KdV equation 
(\ref{kdv}). As higher orders are reached, this $u_0$ is required 
to satisfy also the higher order equations of the KdV hierarchy, 
which amounts to include dependences on the
higher order times $\tau_5$, $\tau_7$, etc. However, there is 
an important difference: the secular-producing term in each 
order of the perturbative scheme will come not only from the
linear term, but from both the linear and the nonlinear terms. 
As a consequence, the slow time normalizations obtained from 
the linear dispersion relation expansion will not be able to 
remove the secular-producing terms in this case. In other words, 
new slow time normalizations will be 
needed to get a secularity-free perturbative series. These new 
normalizations can be easily found by properly choosing the 
free-parameters left at each order of the perturbation 
scheme~\cite{lecce}. After doing that, we obtain 
the following perturbative series for $u$:
\be
u = \epsilon^2 \left[ 1 - 8 k^2 + 64 k^4 - \cdots \right] 
{\rm tanh}^2 \left[ k x - k \left(1 - 8 k^2 + 64 k^4 - \cdots 
\right) \right] \, .
\ee
Like in the previous case, these series can be summed, 
resulting in
\be
u = \frac{2 k^2}{1 + 8 k^2} \, {\rm tanh}^2 \left[k \left(x - 
\frac{t}{1 + 8 k^2} \right) \right] \; ,
\ee
which is the solution (\ref{bbmtan}) of the RLW equation. As 
already said, however, a new slow time normalization is needed 
in this case to get a secularity-free perturbative series,
which is different from that obtained from the dispersion 
relation expansion.

\section{Study on the Applicability of the Multiple Scale Method}

The multiple scale method is not always able to remove all 
the secular-producing terms of a perturbative series~\cite{komi}. 
In some cases, nonintegrable effects may preclude the existence 
of uniform asymptotic expansions. Considering that the RLW is 
nonintegrable, the purpose of this section will be to make a 
brief discussion on how those effects appear in the 
higher order terms of the perturbative series for the specific 
case of the RLW equation. The approach we are going to use is 
that developed by Kodama and Mikhailov~\cite{komi}.

Let us start by defining slow variables according to
\be
u = \epsilon v \quad ; \quad \xi = \epsilon^{1/2} (x-t) 
\quad ; \quad \tau_3 = \epsilon^{3/2} t \; .
\ee
In these new coordinates, and up to terms of order $\epsilon^2$, 
Eq.(\ref{bbm}) becomes
\be
v_{\tau_3} = \partial_{\xi} \left[ 3 v^2 - v_{\xi\xi} + \epsilon 
\partial_{\xi\xi} \left(3 v^2 - v_{\xi\xi} \right) + \epsilon^3 
\partial_{(4\xi)} \left( 3 v^2 - v_{\xi\xi} \right) + \cdots
\right] \; .
\label{62}
\ee
Then, we make a near identity transformation~\cite{ko} given by
\be
v = w + \epsilon \Phi(w) + \epsilon^2 \Psi(w) + {\cal O}
(\epsilon^3) \; ,
\label{63}
\ee
where, by reasons of scaling-weight invariance, the differential 
polinomials $\Phi$ and $\Psi$, which are allowed to be nonlocal, 
can involve only the following terms:
\be
\Phi = \alpha w^2 + \beta w_{\xi\xi} + \gamma w_{\xi} \partial^{-1} 
w \; ,
\label{64}
\ee
\begin{eqnarray}
\Psi = a w^3 &+& b (w_\xi)^2 + c w w_{\xi\xi} + d w_{(4\xi)} + 
e w w_{\xi} \partial^{-1} w 
\nonumber \\
&+& f w_\xi \partial^{-1} (w^2) 
+ g w_{\xi\xi\xi} \partial^{-1} w + h w_{\xi\xi} 
(\partial^{-1} w)^2 \; .
\label{65}
\end{eqnarray}
Substituting into (\ref{62}), we obtain
\be
w_{\tau_3} = K_3 + \epsilon K_5 + \epsilon^2 K_7 + \cdots \; ,
\ee
with
\be
K_3 = \partial_\xi M_0 \; ,
\ee
\be
K_5 = \partial_\xi \left(M_1 + \partial_{\xi\xi} M_0 \right) - 
\frac{\delta \Phi}{\delta w} 
\left(\partial_\xi M_0 \right) \; ,
\ee
\begin{eqnarray}
K_7 = \partial_\xi \left( M_2 + \partial_{\xi\xi} M_1 + 
\partial_{(4\xi)} M_0 \right) &-& \frac{\delta \Psi}{\delta w} 
\left(\partial_\xi M_0 \right) \nonumber \\
&-& \frac{\delta \Phi}{\delta w} 
\left[\partial_\xi \left(M_1 + \partial_{\xi\xi} M_0 \right) - 
\frac{\delta \Phi}{\delta w} 
\left(\partial_\xi M_0 \right) \right] \; ,
\end{eqnarray}
where we have introduced the notation:
\be
M_0 = 3 w^2 - w_{\xi\xi} \; ,
\ee
\be
M_1 = 6 w \Phi - \Phi_{\xi\xi} \; ,
\ee
\be
M_2 = 3 \Phi^2 + 6 w \Psi - \Psi_{\xi\xi} \; .
\ee

At order $\epsilon^0$ we find
\be
K_3 = F_3 \equiv 6 w w_{\xi} - w_{\xi\xi\xi} \; ,
\ee
that is, $K_3$ is the symmetry of order $\epsilon^0$ of the 
KdV equation. At the next order, by
properly choosing $\alpha$, $\beta$ and $\gamma$, we find
\be
K_5 = F_5 \; ,
\ee
with $F_5$ defined by Eq.(kdv5).This means that there exists 
a near-identity transformation (\ref{63})-(\ref{65}) such 
that $K_5$ is the symmetry of order $\epsilon$ of the KdV 
equation. In the first two orders, therefore, no
problems appear. This is a general result that holds for any 
equation, not only for the particular case of the RLW equation. 
It is in the next order that the so called 
obstacles~\cite{komi} show up. In fact, in the next order we get
\be
K_7 = F_7 + O(w) \; ,
\ee
with $F_7$ defined by Eq.(\ref{kdv7}), and $O(w)$ representing 
the obstacle, which is given by
\begin{eqnarray}
O(w) &=& \left(- \frac{32}{3} - 3 g \right) w w_{(5\xi)} + 
\left( - \frac{20}{3} - 3 c
- 24 d - 3 g \right) w_\xi w_{(4\xi)} \nonumber \\ 
&+& \left( - \frac{508}{3} + 6 a + 2 f - 18 g \right)
w^3 w_{\xi} + \left( 22 - 3 c - 6 b - 60 d \right) w_{\xi\xi} 
w_{\xi\xi\xi} \nonumber \\ 
&+& \left( \frac{700}{3} - 18 a - 12 c - 6 f + 72 g \right) 
w w_{\xi} w_{\xi\xi} + \left( \frac{224}{3} - 3 f + 21 g 
\right) w^2 w_{\xi\xi\xi} \nonumber \\
&+&  \left(\frac{158}{3} - 6 a - 6 b - 3 f + 18 g \right) 
(w_{\xi})^3 \; .
\end{eqnarray}
The important point is that, for an arbitrary KdV hierarchy 
solution $w$, it is not possible to choose $a, b, \dots , g$ 
in such a way to have a vanishing obstacle. However, as an 
explicit calculation easily shows, when $w$ is a solitary-wave 
solution of the KdV hierarchy, there is a
near-identity transformation leading to $O(w)=0$.

The above considerations are important in the sense that 
they clarify the results obtained in the previous sections 
concerning the solitary-wave related secularities. But, at the 
same time, they put in evidence the limitations of the 
perturbative scheme which, as we now know, can not be extended to 
the two-or-more soliton solutions in the non-integrable case. On 
the other hand, for integrable systems, like for example the 
shallow water wave equation, the multiple scale method will be 
able to handle both, the solitary-wave and the N-soliton related 
secularities~\cite{sww} since no obstacles will be present in 
either case.

\section{Final Remarks}

We have applied a multiple-time version of the reductive 
perturbation method to study the solitary-wave solution of 
the RLW equation. As it has already been shown~\cite{nois1}, 
the use of multiple time-scales allows for the elimination 
of all solitary-wave related secular-producing terms appearing 
in the evolution equations of the higher-order terms of the 
wave-field. Moreover, it has also been shown~\cite{lecce} 
that these secularities are automatically removed if the slow 
time-scales are normalized according 
to the long-wave expansion of the dispersion relation of the 
original equation. By using this strategy, we have succeeded 
in expressing the solitary-wave solution of the RLW equation 
as a sum of solitary-waves satisfying simultaneously, in the 
slow coordinates, all equations of the KdV hierarchy. Similar 
results have been shown to hold also for the 
Boussinesq~\cite{nois2} and the shallow water wave~\cite{sww} 
equations. However, while in these two cases the 
solitary-wave solution was obtained due to a truncation of 
the perturbative series, the RLW solitary-wave was obtained 
by summing the perturbative series. 

To finish, let us make the following considerations. If we assume 
the RLW equation to be an exact model equation, as we have in 
fact done, the KdV equation appears as its long wave leading 
order approximation. This is one more confirmation of the 
widely known property of the KdV equation, which states that 
it holds a unique, privileged and universal meaning in the 
sense it appears as the leading order approximation of any 
weakly nonlinear dispersive systems, as for example that 
represented by the RLW equation. From this point of view, 
the old dispute~\cite{krus} on the equivalence of the RLW 
and the KdV equations would be made on a different ground since 
the RLW equation should be compared not to the KdV equation, 
but to the whole set of equations of the KdV hierarchy. In other 
words, the RLW equation should be compared not to its leading 
order approximation, but to the whole perturbative series. 
And according to our results, as far as solitary-waves 
are concerned, the RLW equation is indeed equivalent to the KdV 
hierarchy since a solitary-wave of the RLW equation is nothing 
but an infinite series given by the sum of solitary-waves 
satisfying simultaneously all equations of the KdV hierarchy, 
each one in a different slow time variable.

\vspace{1 cm}
\section*{Acknowledgements}

The authors would like to thank J. L\'eon for useful discussions. 
They would also like to thank Y. Kodama and A. V. Mikhailov for 
sending the preprint of reference [8] to them. This work has been 
partially supported by CNPq (Brazil) and CNRS (France).


\begin{thebibliography}{999}
\bibitem{pere}
D. H. Peregrine, J. Fluid Mech. {\bf 25}, 321, (1966).
\bibitem{bbm}
T. B. Benjamin, J. L. Bona and J. J. Mahony, Philos. Trans. Roy. 
Soc. London A {\bf 272}, 47 (1972).
\bibitem{akns}
M. J. Ablowitz, D. J. Kaup, A. C. Newell and H. Segur, Stud. 
Appl. Math. {\bf 53}, 249 (1974).
\bibitem{koda1}
Y. Kodama, in {\it Nonlinear Water Waves, IUTAM Symposium}, 
Tokyo, Japan, 1987, ed. by K. Horikawa and H. Maruo 
(Springer-Verlag, Berlin, 1988).
\bibitem{nois1} 
R. A. Kraenkel, M. A. Manna and J. G. Pereira, J. Math. Phys. 
{\bf 36}, 307 (1995).
\bibitem{lecce} 
R. A. Kraenkel, M. A. Manna, J. C. Montero and J. G. Pereira, 
Proceedings of the Workshop {\it Nonlinear Physics: Theory and 
Experiment}, Gallipoli, Lecce, 1995 (World Scientific, Singapore, 
1995) (patt-sol/9509003).
\bibitem{kota}
Y. Kodama and T. Taniuti, J. Phys. Soc. Jpn. {\bf 45}, 298 (1978).
\bibitem{komi}
Y. Kodama and A. V. Mikhailov, {\it Obstacles to Asymptotic 
Integrability}, Preprint 1995
(to appear in a memorial volume dedicated to Irene Dorfman).
\bibitem{ko}
Y. Kodama, Phys. Lett.A {\bf 107}, 245 (1985).
\bibitem{nois2}
R. A. Kraenkel, M. A. Manna, J. C. Montero and J. G. Pereira, 
J. Math. Phys. {\bf 36}, 6822 (1995).
\bibitem{sww}
R. A. Kraenkel, M. A. Manna, J. C. Montero and J. G. Pereira, 
{\it The N-Soliton Dynamics of the Shallow Water Wave Equation 
and the Korteweg-de Vries Hierarchy},
Preprint IFT-P.040/95 (patt-sol/9509001).
\bibitem{krus}
M. Kruskal, in {\it Dynamical Systems, Theory and Applications}, 
Lecture Notes in Physics 38, ed. by J. Moser (Springer-Verlag, 
Berlin, 1975).

\end{thebibliography}
\end{document}